\documentclass[runningheads]{svmult}

\usepackage{makeidx}   
\usepackage{graphicx}  
\usepackage{subeqnar}  
\usepackage{multicol}  
\usepackage{physprbb}  
\makeindex             

\begin{document}
\title*{Light Curves of Type Ia Supernovae as a Probe for an Explosion Model}
\toctitle{Light Curves of Type Ia Supernovae 
\protect\newline as a Probe for an Explosion Model}
%
%
\titlerunning{Light Curves of SNe Ia}
%
\author{Elena Sorokina\inst{1}
        \and Sergey Blinnikov\inst{2}}
\authorrunning{Elena Sorokina and Sergey Blinnikov}
%
%
\institute{Sternberg Astronomical Institute, 119992 Moscow, Russia, and MPA, Garching, Germany
\and ITEP, 117218, Moscow, Russia, and MPA, Garching, Germany }

\maketitle              

\begin{abstract}
We present theoretical UBVI- and bolometric light curves of SNe Ia 
for several explosion models, computed with our multi-group 
radiation hydro code.
We employ our new corrected treatment for line opacity in the expanding
medium.
The results are compared with observed light curves. 
Our goal is to find the most viable
thermonuclear SN model that gives good fits not only to a typical SN Ia 
light curve, but also to X-ray observations of young SNIa remnants.
It appears that classical 1D SNIa models, such as deflagration W7 and 
delayed detonation DD4, fit the light curves not so good as a new 3D 
deflagration model by Reinecke et al (which is averaged over angles 
for our LC modelling). 
This model seems good also
in reproducing X-ray observations of Tycho SNR. 
We believe that the main feature of this model which allows us to get correct 
radiation during the first month, as well as after a few hundred years, 
when an SNR forms, is strong mixing that pushes material enriched in iron and 
nickel to the outermost layers of SN ejecta.
\end{abstract}

\section{Introduction}

At the moment, there are many models of thermonuclear explosion of a star, 
that lead to the event we know as a Type~Ia Supernova (SN~Ia). 
Some of them were discussed in the talk 
by J.Niemeyer~\cite{sorokina.Nhere}.
Only a few parameters, such as kinetic energy and total $^{56}$Ni production, 
can be derived directly from the explosion modelling and compared 
with the observational values.
The subsequent evolution of the exploded star gives us 
much more possibilities to compare models and to decide which one fits
observations better by reproducing more details in SN~Ia light curves and 
spectra.
We will focus here on the broad-band UBVI and bolometric light curve 
computations for SN~Ia models.

There are several effects in SNe physics which lead to difficulties 
in the light curve modelling of any type of SNe.
For instance, an account should be taken correctly for deposition 
of gamma photons produced in decays of radioactive isotopes, mostly $^{56}$Ni
and  $^{56}$Co.
After being emitted, gamma photons travel through the ejecta and can 
finish up in either thermalization or in non-coherent scattering processes.
To find this one has to solve
the transfer equation for gamma photons together with hydrodynamical 
equations.
Full system of equations should involve also radiative transfer equations 
in the range from soft X-rays to infrared for the expanding medium. 
There are millions of spectral lines that form SN spectra, and it is not 
a trivial problem to find a convenient way how to treat them even 
in the static case.
The expansion makes the problem much more difficult to solve: 
hundreds or even thousands of lines give their input into emission and
absorption at each frequency.

On the first glance, 
modelling of SNe~Ia seems easier than
of other types of supernovae, since the hydro part is very simple: 
coasting stage starts very early, there are no shocks, and no
additional heating from them.

On the other hand, much more difficulties arise in the radiation part. 
SNe~Ia becomes almost transparent in continuum at the age of a few weeks.
This means that NLTE effects are stronger than for other types of supernovae.
Radiation is decoupled from
matter within the entire SN~Ia ejecta even before maximum light
(which occurs around the 20th day after explosion), see e.g.
\cite{sorokina.EPThn} or  Fig.~2 in \cite{sorokina.SBring02}.
In this case one cannot ascribe to radiation the gas temperature, or any other
temperature, since SN~Ia spectrum differs strongly from a blackbody one. 
One has to solve a system of time-dependent transfer equations in many energy groups
instead, with an accurate prescription for treatment of a huge number of spectral lines,
which are the main source of opacity in this type of SN \cite{sorokina.BHM,sorokina.PE00}.

\section{Method}

We compute broad-band UBVI and bolometric light curves of SNe~Ia with a
multi-energy radiation hydro code {\sc STELLA}.
Time-dependent equations for the angular moments of intensity  in fixed 
frequency bins are coupled to Lagrangean hydro equations and solved 
implicitly.
Thus, we have no need to ascribe any temperature to the radiation:
the photon energy distribution may be quite arbitrary.

While radiation is nonequilibrium in our approximation, ionization and
atomic level populations are assumed to be in LTE.
The effect of line opacity is treated as an expansion opacity according to
Eastman \& Pinto 1993~\cite{sorokina.EP} and to our new
recipes~\cite{sorokina.SBring02}.
We have compared the results and found that infrared bandpass (in which the
ejecta are the most transparent) is more sensitive to the treatment of
opacity than UBV (see Fig.~4 in \cite{sorokina.SBring02}
for the comparison of SN~Ia light curves
calculated with these two approaches), so one should be very careful 
on this point.
To simulate NLTE effects we used the approximation of the
absorptive opacity in spectral lines.
NLTE results \cite{sorokina.BHM} and ETLA approach \cite{sorokina.PE00} 
demonstrate that fully absorptive lines gives us an acceptable 
approximate description of NLTE effects.

We treat gamma-ray opacity as a pure absorptive one, and solve the
$\gamma$-ray transfer equation in a one-group approximation 
following~\cite{sorokina.SSH}.
The heating by decays  $^{56}$Ni  $\to$ $^{56}$Co$\to$ $^{56}$Fe is
taken into account.

In the calculations of SN~Ia light curves we use up to 200 frequency bins and up
to $\sim 400$ zones as a Lagrangean coordinate.

\section{Light curves of SNe~Ia}

In our previous work~\cite{sorokina.ourLC} we studied the behaviour 
of the broad-band optical and ultraviolet light curves for classical 1D models 
of SNe~Ia: Chandrasekhar-mass models W7 (deflagration)~\cite{sorokina.W7} 
and DD4 (delayed detonation)~\cite{sorokina.DD4}, and sub-Chandrasekhar-mass 
models LA4 (helium detonation; actually, an averaged 2D model)~\cite{sorokina.livne} 
and WD065 (detonation with low $^{56}$Ni production)~\cite{sorokina.pilar}.
Here we concentrate on a new 3D deflagration model 
by M.Reinecke et al. \cite{sorokina.martin} (MR, hereafter).
Working in more dimensions they have to involve less free parameters 
than was needed in 1D, so they get their  model almost from the ``first
principles''.

\begin{table}
\caption{Parameters of SN~Ia models}
\begin{center}
\begin{tabular}{llll}
\hline
 Model & Mass ($M_\odot$) & $E_{\rm kin}$, foe & $M_{{}^{56}{\rm Ni}}\; (M_\odot$) \\
\hline
 W7    & 1.38         & 1.2                & 0.6 \\
 DD4   & 1.38         & 1.23               & 0.6 \\
 MR    & 1.4          & 0.46               & 0.42 \\
\hline
\end{tabular}                               
\end{center}                                
\end{table}

The main features of the 3D model are compared with the ones of classical 
1D models in the Table.
From the first glance it seems that the light curves for the models which 
differ so much could not be similar.

\begin{figure}[ht]  
  \centerline{\includegraphics[width=0.6\textwidth]{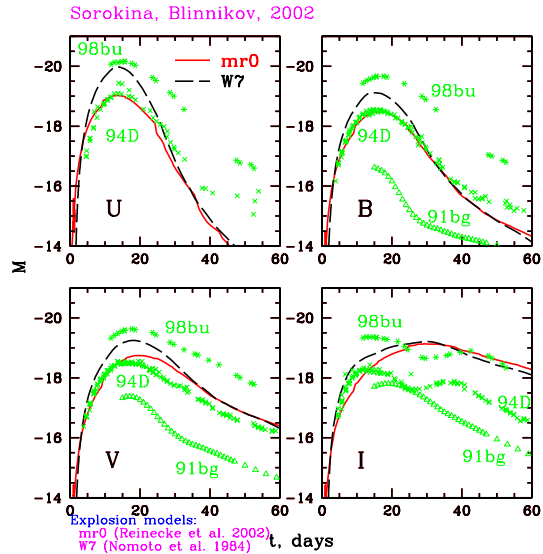} 
}
  \caption{UBVI light curves for the 3D (MR; red solid) and 1D 
(W7; black dashed) models. Crosses, stars and triangles show the light curves
for three observed SNe~Ia.}
  \label{sorokina.figw7mr0}
\end{figure}

\begin{figure}[ht]  
  \centerline{\includegraphics[width=0.45\textwidth]{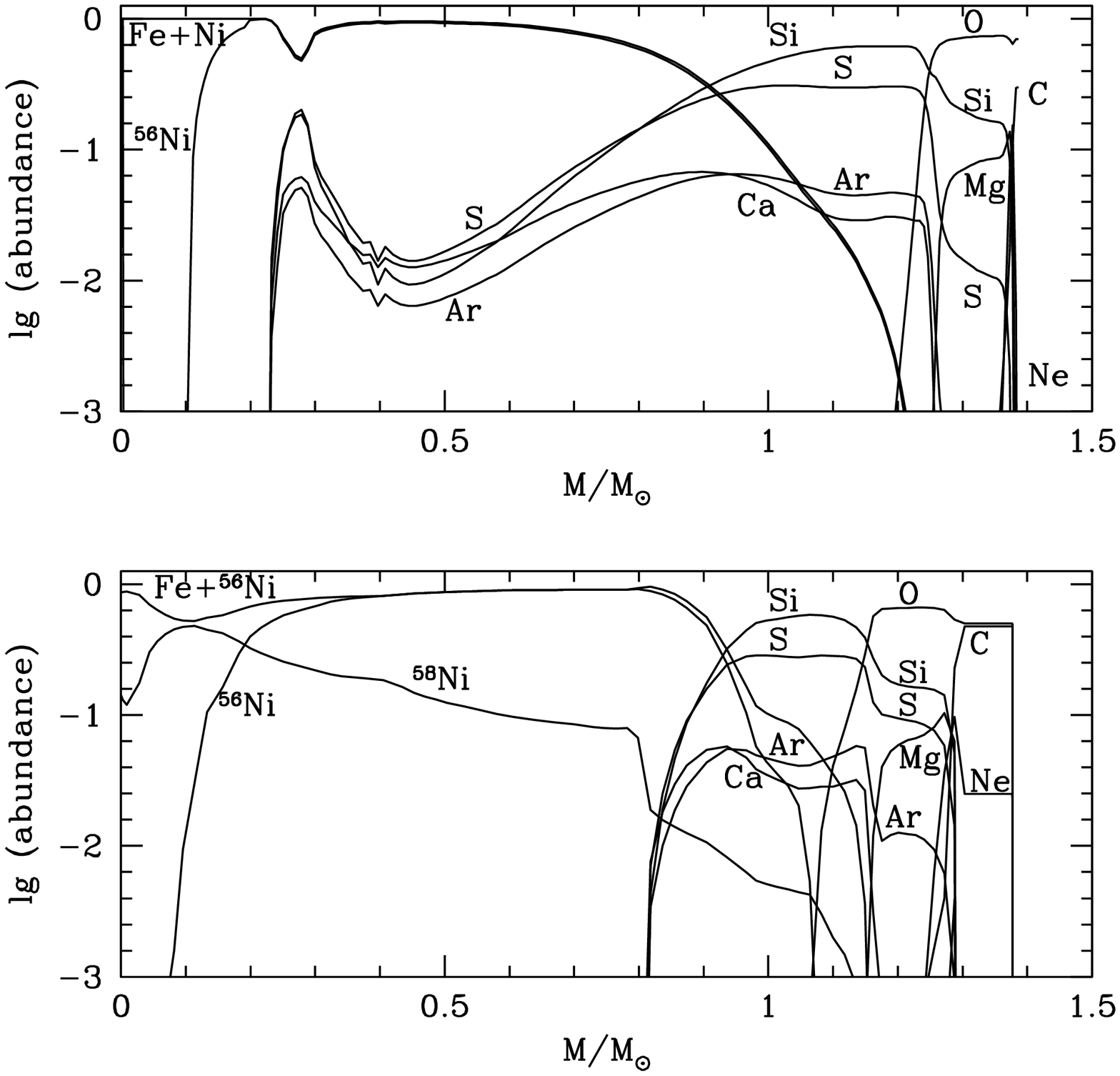} 
              \includegraphics[width=0.45\textwidth]{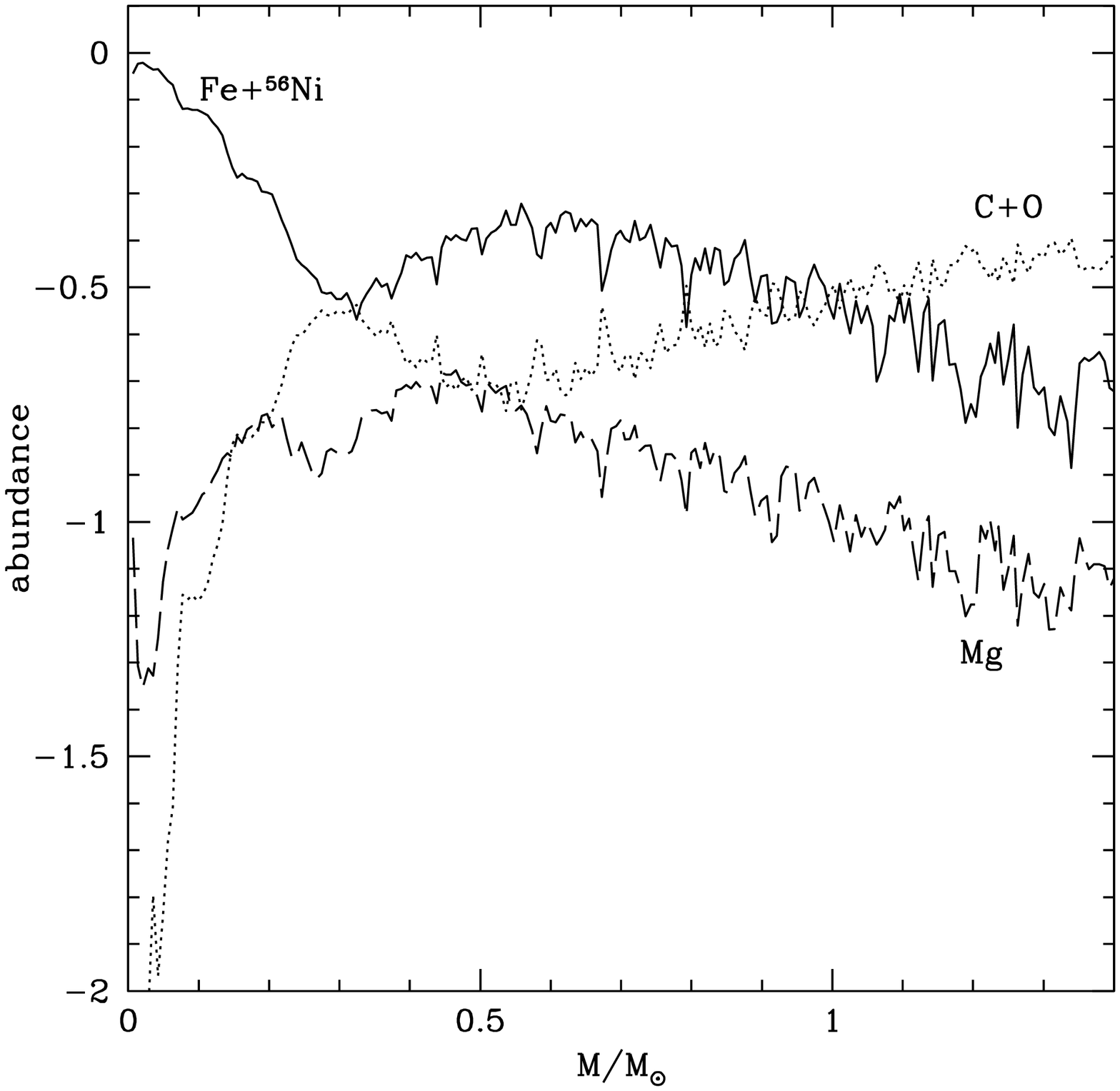}} 
  \caption{Element distributions for DD4 (top left) W7 (bottom left) and MR    
  (right) models.}
  \label{sorokina.figchem}
\end{figure}

\begin{figure}[ht]
  \centerline{\includegraphics[width=0.6\textwidth]{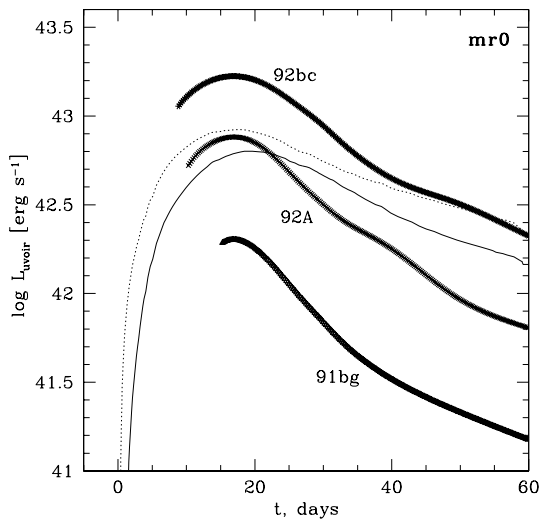}} 
  \caption{Total (dotted) and {\it UVOIR} (solid) bolometric luminosity
of MR model compared with observations \protect\cite{sorokina.CLV} } 
  \label{sorokina.figbol}
\end{figure}

Nevertheless, Fig.~\ref{sorokina.figw7mr0} demonstrates that they are similar 
in many details.
The possible reasons for this can be understood if one has a look 
at the element distribution over the ejecta.
The compositions for W7 and MR models are shown in Fig.~\ref{sorokina.figchem}.
At the moment of our light curve computation the full nuclide  yields for MR 
were not yet obtained.
Therefore, the model consisted of the elements, which were chosen 
as representative examples for the energy release calculation, namely ``Fe''
for iron group elements that were divided onto 80\% of $^{56}$Ni and 20\% 
of $^{56}$Fe, ``Mg'' for intermediate mass elements, and unburned C and O 
in equal proportion.
The instabilities that have developed in the 3D model were not supposed
to be so huge in approximate 1D models of explosion.
This has led to the differences in the nickel distribution over the ejecta: 
it is mixed to the outermost layers in the 3D model.
These layers became much more opaque than in the 1D models, and, 
despite having less than a half of kinetic energy, the 3D model has a
photospheric velocity comparable to that in the 1D models.
It is probably still a bit too low, so the light curve is wider, since the ejecta
expand a bit slower, and photons are locked inside them for a bit longer time.
The broad-band light curves for MR model fit the observations 
of one of the typical SN~Ia, SN~1994D, in $U$ and $B$ bands surprisingly well, 
while classical 1D models, such as  W7 and  DD4, show faster decline 
in the optics than it is observed.
Unfortunately, the bolometric light curve for MR model 
(Fig.~\ref{sorokina.figbol}) is somewhat too slow.
The ejecta must expand with higher speed to let photons to diffuse out
faster.

\begin{figure}[ht]
  \centerline{\includegraphics[width=0.6\textwidth]{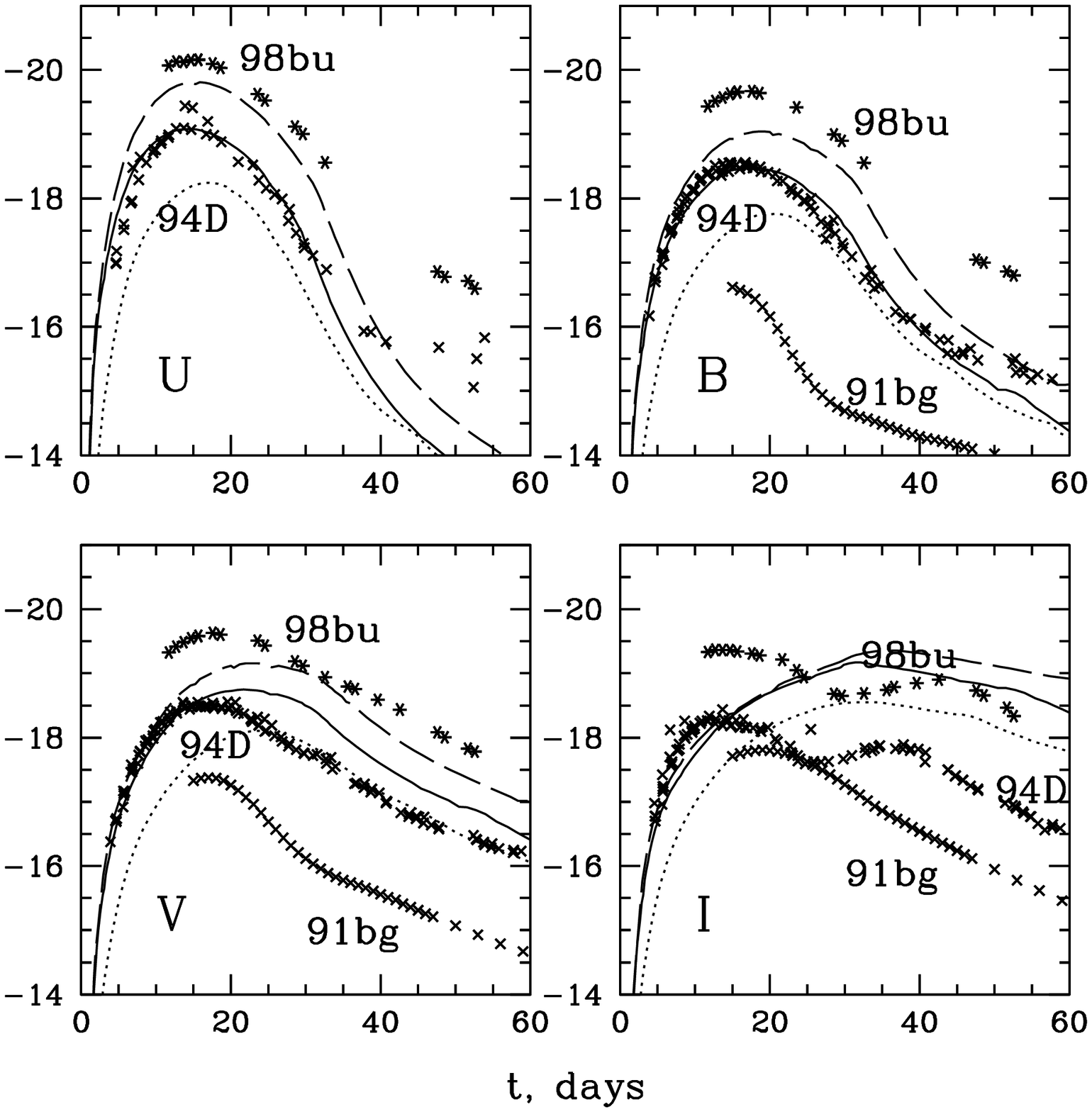}} 
  \caption{UBVI light curves of 3D MR model averaged over different opening
angles. Solid line for the whole $4\pi$; dashed for a $14^\circ$ cone 
with maximum $^{56}$Ni abundance; dotted for a $14^\circ$ cone with minimum
$^{56}$Ni abundance.} 
  \label{sorokina.fig3ni}
\end{figure}

Since our light curve code is 1D, we cannot model 3D effects  directly.
We tried  to estimate them by averaging the model over different solid angles.
The main question we wished to answer was: how different can the SN~Ia 
look like for an observer from different sides.
To start with, we just compared the models averaged over cones 
with opening angles of $14^\circ$ with the one averaged 
over the whole $4\pi$~\cite{sorokina.MR}.
We tried to choose two limiting cases: the angles with minimal and maximal 
nickel abundances, i.e. one with nickel bubble extended to the surface, 
and another one between such bubbles, with nickel only in the center.
Since total $^{56}$Ni mass differs in these models, the light curves are 
also very different (Fig.~\ref{sorokina.fig3ni}).
They just simulate different models, but not a model observed 
from different sides.

\begin{figure}[ht]
  \centerline{\includegraphics[width=0.6\textwidth]{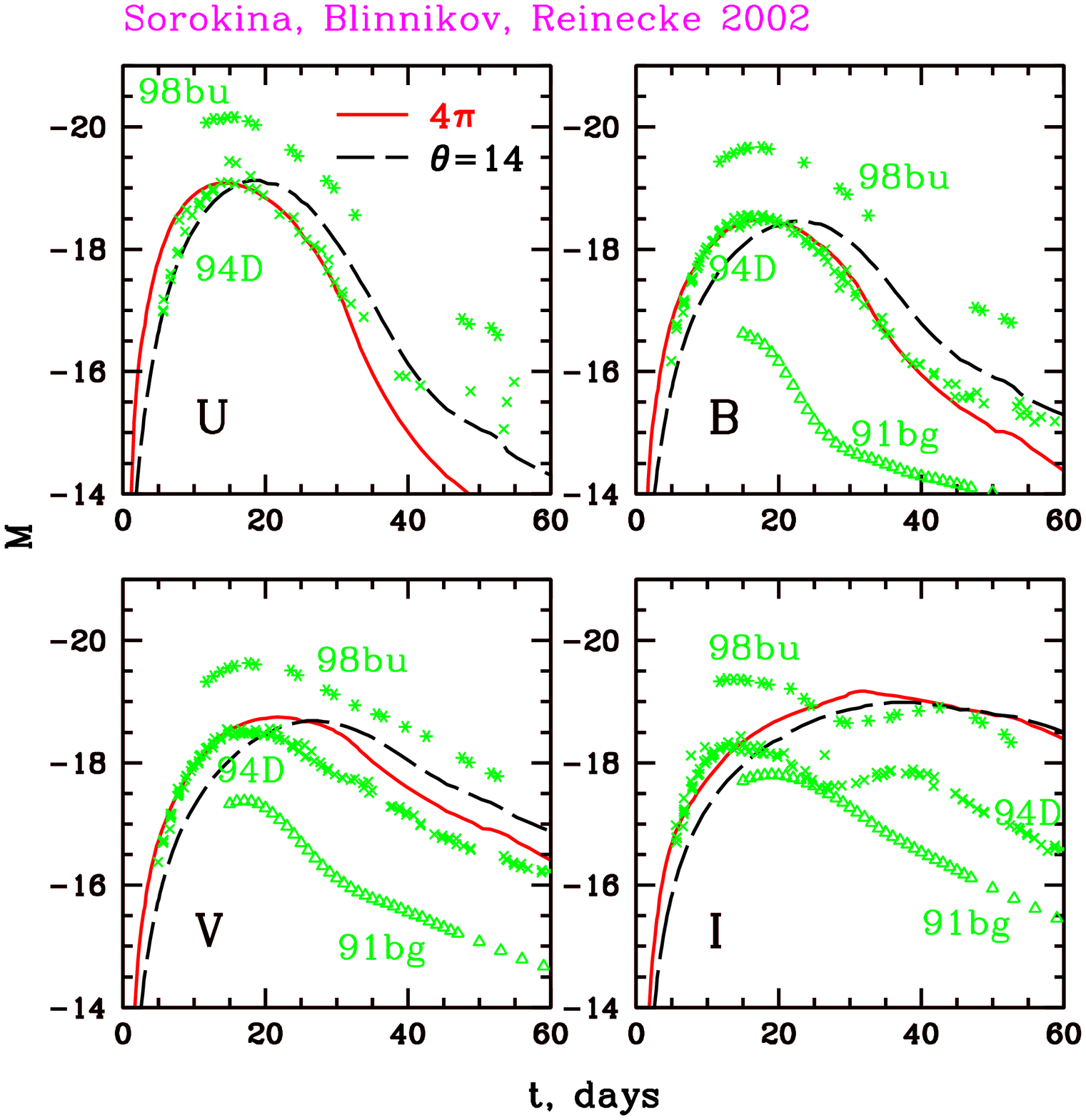}} 
  \caption{Comparison of light curves of the MR model averaged over $4\pi$ 
(red solid); and for a $14^\circ$ cone with $^{56}$Ni abundance rescaled 
to preserve the total  $^{56}$Ni mass (black dashed).}
  \label{sorokina.figmr1ni}
\end{figure}

In reality, the total energetics one observes is defined by total 
$^{56}$Ni mass. 
So we have to preserve it when averaging the model over different directions.
In the next experiment we have taken the same solid angle with low $^{56}$Ni
abundance and enhanced the mass of $^{56}$Ni in every mass zone 
by the same ratio, so that the total $^{56}$Ni mass became equal to the one 
of the original 3D model.
The resulting light curve is compared with the fully averaged case 
in the fig.~\ref{sorokina.figmr1ni}.
Since $^{56}$Ni mass fixes the brightness of SNe~Ia~\cite{sorokina.arnett},
the light curves become equally bright, but the maximum is shifted in time
due to a bit different energetics and mixing.
We believe that differences between 1D and 3D light curve calculations 
should not be much stronger 
than the difference between these two light curves,
when other physical assumptions are fixed.

There is also another reason which allows us to believe that the MR model 
is better than the classical 1D models.
Tycho SNR is believed to be the remnant of SN~Ia.
We calculated in details the X-ray emission of Tycho SNR, allowing for time-dependent 
ionization and recombination, and compared the computed X-ray spectra and 
images in narrow filter bands with XMM-Newton observations of the Tycho.
Our preliminary results \cite{sorokina.tycho} show
that all Chandrasekhar-mass models produce
similar X-ray spectra at the age of Tycho, but they differ strongly 
in the predictions of how the remnant should look like in the lines 
of different ions due to very different distribution of elements in the ejecta 
for 1D and 3D models.
We found that W7 and DD4 models 
produce rather wide ring in Fe lines, while it is narrow for MR model.
The image for the latter model is very similar to what is observed.

\section{Conclusions}

The new 3D SN~Ia model MR \cite{sorokina.martin} is very appealing.
Yet it is not a final one: a detailed post-processing of nucleosynthesis 
changes the composition.
It has been  done 
very recently  \cite{sorokina.claudia}, and it is not yet
checked in the light curve calculation.
Our light curve computations are also preliminary, since
more work is needed on the expansion opacity.
Hopefully, none of the required improvements will spoil the light curve 
of this model and its X-ray spectra on the SNR stage, 
since the specific qualities of the model can be primarily explained 
by the enrichment of the outermost layers of SN ejecta by  Fe and Ni.

The SN light curve modelling still has a  lot 
of physics to be added, such as a 3D time-dependent 
radiative transfer, including as much as possible of NLTE effects, which are 
especially essential for SNe~Ia  \cite{sorokina.hoeflich}.
All this will improve our understanding of thermonuclear supernovae.

\subsection*{Acknowledgements}

The authors are grateful to Wolfgang Hillebrandt, 
to the organizers of the conference and to the MPA staff 
for all their help, to Stan Woosley for his hospitality at UCSC
where a major part of the work on expansion opacity was done, and to Bruno
Leibundgut for providing us with the data \cite{sorokina.CLV} in
electronic form.

The work is supported in Russia by RFBR grant 02-02-16500, 
in the US, by NASA grant NAG5-8128, and in Germany, by MPA visitor program.

\end{document}